\documentstyle[twocolumn,prl,aps]{revtex}
\begin{document}
\title
{Skyrmion Excitations in Quantum Hall Systems}

\author
{X.C. Xie}

\address
{
Department of Physics, Oklahoma State University, Stillwater, OK 74078}

\author
{Song He}

\address
{AT\&T Bell Laboratories, Murray Hill, NJ 07974}

\address{\rm (Draft on \today )}
\address{\mbox{ }}
\address{\parbox{14cm}{\rm \mbox{ }\mbox{ }
Using finite size calculations on the surface of a sphere
we study the topological (skyrmion) excitation in
quantum Hall system with spin
degree of freedom at filling factors around $\nu=1$.
In the absence of Zeeman energy,
we find, in systems with one quasi-particle or one quasi-hole, the lowest
energy band consists of states with $L=S$, where $L$ and $S$ are the
total orbital and spin angular momentum.
These different spin states are almost degenerate in the thermodynamic limit
and their symmetry-breaking ground state
is the state with one skyrmion of infinite
size. In the presence of Zeeman energy, the skyrmion size is determined by the
interplay of the Zeeman energy and electron-electron
interaction and
the skyrmion shrinks to a
spin texture of finite size.
We have calculated the energy gap of the system at
infinite wave vector limit as a
function of the Zeeman energy and find
there are kinks in the energy gap associated
with the shrinking of the size of the skyrmion.
}}
\address{\mbox{ }}
\address{\parbox{14cm}{\rm PACS No: 05.30.-d, 73.50.Jt, 74.20.-z}}
\maketitle

\makeatletter
\global\@specialpagefalse
\def\@oddhead{REV\TeX{} 3.0\hfill Preprint, 1995}
\let\@evenhead\@oddhead
\makeatother

  The properties of two-dimensional electron systems exhibit rich
physics in the presence of strong magnetic field. Electron-electron
Coulomb interactions add more interesting many body phenomena to the systems,
such as fractional quantum Hall effect (FQHE)\cite{tsui82,laughlin83}.
Over the last few years, both experimental\cite{exp} and
theoretical\cite{theory} studies of quantum Hall effect
have been extended to multi-component
systems, such as systems where electron spins play important roles
or double quantum well systems. Most recently, a new type of topological
excitation-{\it skyrmion} is predicted to occur around filling
factor $\nu =1/m$ with odd integer $m$
by field theory method\cite{lee90,sondhi93,moon} and
Hartree-Fock calculation\cite{fertig94}. The physical consequences of the
existence of
skyrmions or spin texture excitations may have already been
observed in a recent NMR experiment.\cite{barrett95}

In this paper, using finite size calculations in the spherical geometry,
we study the skyrmion excitations in
quantum Hall systems with spin degree of freedom at filling factors around
$\nu=1$.
In the absence of Zeeman energy,
we find, in systems with one quasi-particle or one quasi-hole, the lowest
energy band consists of states with $L=S$, where $L$ and $S$ are the
total orbital and spin angular momentum.
We show that the existence of this low energy band with $L=S$ in the
microscopic
excitation spectra is a necessary condition for the existence of
skyrmion excitations.
These low energy states are almost degenerate in the thermodynamic limit
and their symmetry-breaking ground state
is the state with one skyrmion of infinite
size. In the presence of Zeeman energy, the skyrmion size is determined by the
interplay of the Zeeman energy and electron-electron
interaction and
the skyrmion shrinks to a
spin texture of finite size.
We have calculated the energy gap of the system at
infinite wave vector limit as a
function of the Zeeman energy and find
there are kinks in the energy gap associated
with the shrinking of the size of the skyrmion.

Our numerical diagonalization is done in the spherical geometry introduced by
Haldane\cite{prange}. In this geometry electrons are placed on the
surface of a sphere under the influence of an uniform radial magnetic
field. The magnetic field is produced by a magnetic monopole of
suitable strength placed at the center of the sphere.
For a monopole of total magnetic flux $N_{\phi}=2q$,
the lowest Landau level (LL) degeneracy
is also $2q+1$.
All single particle states in the lowest LL have fixed angular
momentum value $q$.

At filling factor $\nu=1$, corresponding to electron number
$N=2q+1$, the ground state is spin polarized,
independent of the strength of Zeeman splitting. Now let us study the case
when one additional flux quantum is added or removed from the system.
In Fig.1 we show the energy spectra versus total orbital
angular momentum $L$ for $N=10$ with Coulomb
interactions in the absence of Zeeman splitting.
All possible spin
states from $S=0$ to $S=5$ are plotted in Fig.1. Fig.1(a) is for the
case of one extra flux with $N_{\phi}=N+1$ and Fig.1(b) is for one less
flux with $N_{\phi}=N-1$. The ground state in both cases
has quantum numbers $L=S=0$\cite{sondhi93,moon,rezayi}.
There are two important features in Fig.1: (1) There exists an energy gap
separating the lowest energy band from high energy states;
(2) The lowest energy band consists of states of $L=S$ with all
possible values of $S$ in the range $0\leq S \leq N$.
If short range interaction\cite{prange}
where only the $V_0$ component of its Haldane pseudopotential
is non-zero,
then the states in the lowest energy band are degenerate.
For the Coulomb interaction,
the energy of a state in this band goes up like $L^2$
in the small $L$ limit. The state with $L=S=N/2$ has the
highest energy in the band. Its energy is higher than that of the ground
state by a finite amount of $\Delta_p$. As $N\rightarrow\infty$,
these states become almost degenerate. They can linear combine to
form the macroscopic ground state of the system with
one skyrmion of infinite size.
This macroscopic ground state breaks
the both $L$ and $S$ symmetries.
In the following we will further
demonstrate that the condition $L=S$ is a manifestation
of a classical skyrmion excitations in a quantum system.

\begin{figure}
 \vbox to 5.0cm {\vss\hbox to 6.5cm
 {\hss\
   {\includegraphics{fig1a.ps}
   }
  \hss}
 }
\vspace{2mm}
 \vbox to 5.0cm {\vss\hbox to 6.5cm
 {\hss\
   {\includegraphics{fig1b.ps}
   }
  \hss}
 }
\caption{
Energy spectra versus total orbital angular
momentum $L$ for $N=10$ electrons and all possible spins
$S=0-5$. Energy is in the unit of $e^{2}/\epsilon {\it l}$, where ${\it l}$ is
the magnetic length. (a) Magnetic flux $N_{\phi}=2q+1=11$;
(b) Magnetic flux $N_{\phi}=2q+1=9$.
\label{Fig1}}
\end{figure}

If the boundary conditions are chosen such that the spin directions at north
and south poles are in their respectively
radial directions, then skyrmion states
can be written as
\begin{equation}
{\hat \phi} ({\hat \Omega})=\cos~(g(\theta)-\theta) {\hat e}_{r}
+\sin ~(g(\theta)-\theta) {\hat e}_{\theta},
\end{equation}
where $g(\theta)$ is a smooth and monotonical function satisfying the boundary
conditions $g(0)=0$ and $g(\pi)=\pi$.
The size of a skyrmion is determined by the function $g(\theta)$.
$g(\theta)=\theta$ describes the hedgehog skyrmion with spin in
the radial direction $\hat{r}$.
The spherical skyrmion state in Eq.(1) is equivalent to the more familiar
one in the planar geometry\cite{wilczek}
\begin{equation}
\phi ({\hat r})=(\hat{r} \sin(f(r)),\cos(f(r))),
\end{equation}
where $f(r)$ is a function varying smoothly and monotonically decreasing from
$f(0)=\pi $ to $f(\infty)=0$ as $r$ increases. The connection
between the skyrmion states in the two geometries is through the equation
\begin{equation}
g(\theta)=f(2cotan({\theta}/2))
\end{equation}
We should point out that Eq.(1) is for the skyrmion state with the constraint
$\mid {\hat \phi} ({\hat \Omega}) \mid =1$. If this constraint is relaxed,
the generalized skyrmion state can be defined as
\begin{equation}
{\hat \phi} ({\hat \Omega})=h_{r}(\theta) {\hat e}_{r}
+h_{\theta}(\theta) {\hat e}_{\theta},
\end{equation}
where $h_{r}(\theta)$ does not change sign for $0 \leq \theta \leq \pi$
and the boundary conditions require $h_{\theta} (0)=h_{\theta} (\pi)=0$.
It turns out that this generalized skyrmion state is needed to write
down the quantum many-body skyrmion states as we will discuss later.

{}From Eq.(1) one can write down the basis vectors in the spin space
(${\hat s}_{r}, {\hat s}_{\theta}, {\hat s}_{\phi}$)
in terms of the basis vectors in
real space (${\hat e}_{r}, {\hat e}_{\theta}, {\hat e}_{\phi}$)
as following:
\begin{equation}
\begin{array}{l}
{\hat s}_{r}=\cos~(g(\theta)-\theta) {\hat e}_{r}
+\sin ~(g(\theta)-\theta) {\hat e}_{\theta} \\
{\hat s}_{\theta}=-\sin ~(g(\theta)-\theta) {\hat e}_{r}
+\cos~(g(\theta)-\theta) {\hat e}_{\theta} \\
{\hat s}_{\phi}={\hat e}_{\phi}
\end{array}
\end{equation}
Therefore, a rotation in real space is equivalent as another rotation
in spin space. Now let us look at the quantum mechanical consequences of
the above statement. If we lebel $|\Phi _{sky}>$ as a quantum state for
skyrmion, then a rotation in ${\alpha}$ direction in real space
is identical as another rotation ${\beta}$ in spin spapce,
\begin{equation}
|\Phi _{sky}>= e^{-{\hat \beta}\cdot {\vec S}} e^{-{\hat \alpha}\cdot {\vec L}}
|\Phi _{sky}>.
\end{equation}
where ${\vec L}$ (or ${\vec S}$ is the total orbital (or spin) angular
momentum operator. Since ${\alpha}$ can be in any direction, one must have
$L=S$ where $L$ and $S$ are the eigenvalues of ${\hat L}$ and ${\hat S}$.
Thus, the lowest bands in Fig.1 are quantum manifestations of classical
skyrmion states. In the thermodynamic limit, the lowest band is almost
degenerate, their linear combination gives rise to a skyrmion state.
In this sense, the skyrmion state is spontaneous symmetry-breaking state,
where ${\vec S}$ and ${\vec L}$ symmetries are
broken in the thermodynamic limit.
To see this point further, we can study the operator
\begin{equation}
{\hat \sigma} \cdot {\hat s}_{r}=
\left (
\begin{array}{cc}
\cos (g(\theta)) & \sin (g(\theta)) e^{-i \phi} \\
\sin (g(\theta)) e^{i \phi} & \cos (g(\theta)) \end{array}
\right ),
\end{equation}
and its eigenvector is
\begin{equation}
\xi=\left( \begin{array}{c} \sin(g(\theta)/2)e^{-i\phi /2} \\
-\cos(g(\theta)/2)e^{i\phi /2} \end{array} \right).
\end{equation}
If we label
$(u, v)=(\cos(\theta/2) e^{i \phi/2}, \sin (\theta/2) e^{-i \phi}/2)$,
Then the quantum state for the classical skyrmion
${\hat \phi}({\hat \Omega})$ can be written as
\begin{equation}
\vert \psi \rangle
=C\cdot {\hat P} \prod_{k}
\left( \begin{array}{c} \sin(g(\theta_{k})/2)e^{-i\phi_{k} /2} \\
-\cos(g(\theta_{k})/2)e^{i\phi_{k} /2} \end{array} \right)
\vert \psi _{\nu =1}\rangle,
\end{equation}
where $C$ is the normalization constant and ${\hat P}$ is the projection
operator to the lowest Landau level.
\begin{equation}
\vert \psi _{\nu =1}\rangle
=\prod_{i,j} (u_{i}v_{j}-v_{i}u_{j})
\end{equation}
is the spin polarized ground state at $\nu =1$.
The state for the hedgehog skyrmion with $g(\theta)=\theta$
has been given by
Moon {\it et al}\cite{moon}.
The above skyrmion state can also be written as
\begin{equation}
\vert \psi \rangle
=C' \prod_{k}
\left( \begin{array}{c} v_{k} \\
-\alpha u_{k} \end{array} \right)
\vert \psi _{\nu =1}\rangle ,
\end{equation}
where $0 \leq \alpha \leq 1$ is determined from
$g(\theta)$ and it controls the size
of a skyrmion. To see the size dependence on $\alpha$,
one can calculate the spin expectation value
in the state $\xi ^{+}=(v, -\alpha u)^{+}$,
\begin{equation}
\langle \xi \vert {\hat \sigma} \vert \xi \rangle
= h_{r}(\theta) {\hat e}_{r}
+h_{\theta}(\theta) {\hat e}_{\theta},
\end{equation}
where
\begin{equation}
\begin{array}{c} h_{r}(\theta)=-\alpha +(1-\alpha)\cos \theta
(\alpha \cos^{2}(\theta /2)+\sin^{2}(\theta /2)) \\
h_{r}(\theta)=-(1-\alpha)\sin \theta
(\alpha \cos^{2}(\theta /2)+\sin^{2}(\theta /2)) \end{array}.
\end{equation}
This expectation value corresponds to the generalized classical
skyrmion state as in Eq.(4). Naturely, the size of the skyrmion can be
defined by the position where spin direction is perpendicular to
the radial (and spin)
direction at either pole. With this convention, the skyrmion
size is
\begin{equation}
\theta _{0}=2 \arctan \alpha ,
\end{equation}
which equals $\pi /2$ for the hedgehog skyrmion with $\alpha =1$.

{}From the above discussions, one can see that
$\vert \psi \rangle$ describes a many-body state with all electrons
in the lowest Landau level and
its maximum orbital angular momentum $L$
is equal to $N/2$. Furthermore, $\vert \psi \rangle$
is not eigenstate of either ${\vec L}$ or ${\vec S}$.
Thus, it is consistent to consider this skyrmion
state as a linear combination of the states in the lowest energy
band shown in Fig.1.

\begin{figure}
 \vbox to 5.0cm {\vss\hbox to 6.5cm
 {\hss\
   {\includegraphics{fig2a.ps}
   }
  \hss}
 }
\vspace{2mm}
 \vbox to 5.0cm {\vss\hbox to 6.5cm
 {\hss\
   {\includegraphics{fig2b.ps}
   }
  \hss}
 }
\caption{
Energy spectra versus total orbital angular
momentum $L$ for $N=10$ electrons and magnetic flux
$N_{\phi}=2q+1=12$. (a) The total spin of the system is $S=4$;
(b) The total spin of the system is $S=3$.
\label{Fig2}}
\end{figure}

At exactly filling factor $\nu =1$ corresponding
$n=2q+1$, the ground state is ferromagnetic and gives rise the topological
charge\cite{rajaraman} $Q=0$. When one additional flux is added or
removed from the system, the ground state is antiskyrmion or
skyrmion with $Q=\pm 1$. Next, we would like to study the case
when two flux quanta are added to the system. In this case,
the lowest energy band corresponds
to $Q=2$ topological excitations. Our numerical studies are with particle
number $N=10$. At maximum total spin $S=5$, the lowest
energy band consists of two Laughlin quasiholes and is well
understood\cite{he}. Fig.2 shows the energy spectra $S=4,3$.
Again, there is gap separating the low energy bands with high ones.
The basic question is following: Can the low bands here be understood from the
lowest excitations of $Q=1$? The same idea has been applied in the
spin polarized (or spinless) case where one finds that all the low energy bands
can be understood from the combinations of quasihole (or quasiparticle)
excitations if
their fractional statistics are properly incorporated\cite{he}.
Because of the statistics transmutation in the two dimensional systems,
we consider the excitations at $Q=1$ as bosons.
The excess spins for the excitations are $\Delta S=5,4,3,2,1,0$,
where $\Delta S$ is the spin difference from the ferromagnetic state
at $\nu =1$. Fig.2(a) for $S=4$ corresponding to $\Delta S =1$
should be consisted of $\Delta S_{1}=0$ and $\Delta S_{2}=1$ states of
Fig.1(a).
Thus, ${\hat L}={\hat L}_{1}+{\hat L}_{2}$ with $L_{1}=5$ and $L_{2}=4$,
so the possible values for $L$ are from $1$ to $9$ as shown in Fig.2.(a).
Fig.2(b) for $S=3$ (or $\Delta S=2$) consists of two low energy branchs:
(i) $\Delta S_{1}=1$, $\Delta S_{2}=1$ and $L_{1}=4$, $L_{2}=4$;
(ii) $\Delta S_{1}=0$, $\Delta S_{2}=2$ and $L_{1}=5$, $L_{2}=3$.
The second branch has $L=2-8$ as shown in Fig.2(b).
Because of the bosonic nature, the first branch should have $L=0,2,4,6,8$,
however, $L=8$ state is clearly missing in Fig.2(b). Generally speaking,
we have found in almost all cases there are missing low energy states.
{}From the experience of early studies of spinless case,\cite{he}
the missing states there are caused by hard-core boundary constraint
on the anyonic quasiparticles which pushes some states up in energy.
This implies there exists additional physical constraint for the
spin systems which is
elusive to us at the present time.

Next, we would like to address the experimental aspects of
skyrmion excitations, in particular, what is the signature of
skyrmion excitations in terms of the energy gap measurement.\cite{schmeller}
In Fig.3 we show the energy gaps versus Zeeman energy at $\nu =1$
for $N=12$ with sample thickness $b=2 {\it l}$ and $b=4 {\it l}$.
This plot corresponds to realistic experiment with tilted magnetic
field. In the tilted field experiment, by keeping the perpendicular
field constant one can fix the filling factor and changing the total
field will cause the changes in Zeeman energy.
The energy gap is determined by the interplay of the
Zeeman energy and electron-electron interaction. Electron-electron
interaction favors $S=0$ as excitations shown in Fig.1,
while Zeeman energy favors
maximum spin. At large Zeeman energy, all electron spins are aligned
in the same direction. As the Zeeman energy decreases, electron spins
may flip over to give rise to the kink structures in Fig.3.
As shown in Fig.1, for the skyrmion state the spin angular momentum and orbital
angular momentum $L$ (corresponding to wave vector $q$ in the planar
geometry) is related to each other, and later is related to skyrmion
size. Therefore, the kinks in the energy gap for increasing Zeeman energy
are associated with the
shrinkages of the size of a skyrmion.

\begin{figure}
 \vbox to 5.0cm {\vss\hbox to 6.5cm
 {\hss\
   {\includegraphics{fig3.ps}
   }
  \hss}
 }
\caption{
Energy gaps versus Zeeman energy at filling $\nu =1$ for
sample thickness $b=2 {\it l}$  and $b=4 {\it l}$.
Both energies are in the unit of $e^{2}/\epsilon {\it l}$.
\label{Fig3}}
\end{figure}

In conclusion, we have studied skyrmion excitations at $\nu =1$ using finite
size calculation. The low energy excitations consist of states
with $L=S$ which is shown to be the quantum manifestation of the
classical skyrmion state. We also performed the energy gap calculation
as a function of Zeeman energy and found kinks in the energy gap
associated with the shrinking of the skyrmion size.

One of us (X.C. Xie) thanks Professors Y.S. Wu and J.H.H. Perk
for helpful discussions.

\end{document}